\documentclass[conference]{IEEEtran}
\IEEEoverridecommandlockouts

\usepackage{cite}
\usepackage{amsmath,amssymb,amsfonts}
\usepackage{algpseudocode}
\usepackage{algorithm}
\usepackage{graphicx}
\usepackage{textcomp}
\usepackage{xcolor}
\usepackage{multirow}
 \usepackage{hyperref}
 \hypersetup{
     colorlinks=true,
     linkcolor=blue,
     filecolor=blue,
     citecolor = blue,      
     urlcolor=cyan,
     }

\newcommand{\dk}[1]{\textcolor{black}{#1}} 
\usepackage{booktabs}
\def\BibTeX{{\rm B\kern-.05em{\sc i\kern-.025em b}\kern-.08em
    T\kern-.1667em\lower.7ex\hbox{E}\kern-.125emX}}
\begin{document}


\title{GRPO with State Mutations: Improving LLM-Based Hardware Test Plan Generation\\
{\footnotesize 
}
}

\author{\IEEEauthorblockN{Dimple Vijay Kochar\textsuperscript{1}, Nathaniel Pinckney\textsuperscript{2}, Guan-Ting Liu\textsuperscript{3}, Chia-Tung Ho\textsuperscript{4}, Chenhui Deng\textsuperscript{4},\\ Haoxing Ren\textsuperscript{2}, Brucek Khailany\textsuperscript{2} }
\IEEEauthorblockA{\textit{\textsuperscript{1}Electrical Engineering and Computer Science, Massachusetts Institute of Technology, Cambridge, MA} \\
\textit{\textsuperscript{2}NVIDIA Research, Austin, TX}\\
\textit{\textsuperscript{3}NVIDIA Research, Taiwan}\\
\textit{\textsuperscript{4}NVIDIA Research, Santa Clara, CA}\\
Emails: dkochar@mit.edu, npinckney@nvidia.com, dannliu@nvidia.com,\\ chiatungh@nvidia.com, cdeng@nvidia.com, haoxingr@nvidia.com, bkhailany@nvidia.com}}



\maketitle

\begin{abstract}
RTL design often relies heavily on ad-hoc testbench creation early in the design cycle. While large language models (LLMs) show promise for RTL code generation, their ability to reason about hardware specifications and generate targeted test plans remains largely unexplored. We present the first systematic study of LLM reasoning capabilities for RTL verification stimuli generation, establishing a two-stage framework that decomposes test plan generation from testbench execution. Our benchmark reveals that state-of-the-art models, including DeepSeek-R1 and Claude-4.0-Sonnet, achieve only 15.7-21.7\% success rates on generating stimuli that pass golden RTL designs. To improve LLM-generated stimuli, we develop a comprehensive training methodology combining supervised fine-tuning with a novel reinforcement learning approach, GRPO with State Mutation (GRPO-SMu), which enhances exploration by varying input mutations. Our approach leverages a tree-based branching mutation strategy to construct training data comprising equivalent and mutated trees, moving beyond linear mutation approaches to provide rich learning signals. Training on this curated dataset, our 7B parameter model achieves a 33.3\% golden test pass rate and a 13.9\% mutation detection rate, representing a 17.6\% absolute improvement over baseline and outperforming much larger general-purpose models. These results demonstrate that specialized training methodologies can significantly enhance LLM reasoning capabilities for hardware verification tasks, establishing a foundation for automated sub-unit testing in semiconductor design workflows.
\end{abstract}


\section{Introduction}

Hardware verification follows a well-established continuum from system-level behavioral models to functional implementations to final RTL designs~\cite{bergeron2003writing}. Verification workflows at the system and unit levels rely on golden reference models, external checkers, and constrained random generation~\cite{mathur2007design, aharon1991verification, yuan1999modeling} to validate correctness. 
However, such comprehensive test suites, reference implementations, and substantial engineering infrastructure are often impractical to use at the level of individual RTL module or sub-unit~\cite{wile2005comprehensive}, especially early in the design process.
\dk{This underlines the need for effective automated approaches for low-level RTL code verification.}


Recent advances in large language models (LLMs) have shown promise for automated RTL workflows~\cite{pearce2020dave,thakur2024verigen,liu2024rtlcoder,pei2024betterv,liu2025deeprtl, scalertl2025}, with specialized models demonstrating capabilities in code generation and debugging tasks. However, these approaches primarily address post-generation correction or testbench automation rather than proactive verification reasoning within the models.
Past approaches have also utilized LLM-based agents for generating simpler Python functional models for validation~\cite{correctbench, prov}.
However, the effectiveness of such extensive infrastructure frameworks in understanding complicated RTL with complex timing behaviors remains underexplored.

\begin{figure}[t]
  \centering
  \includegraphics[width=0.9\linewidth]{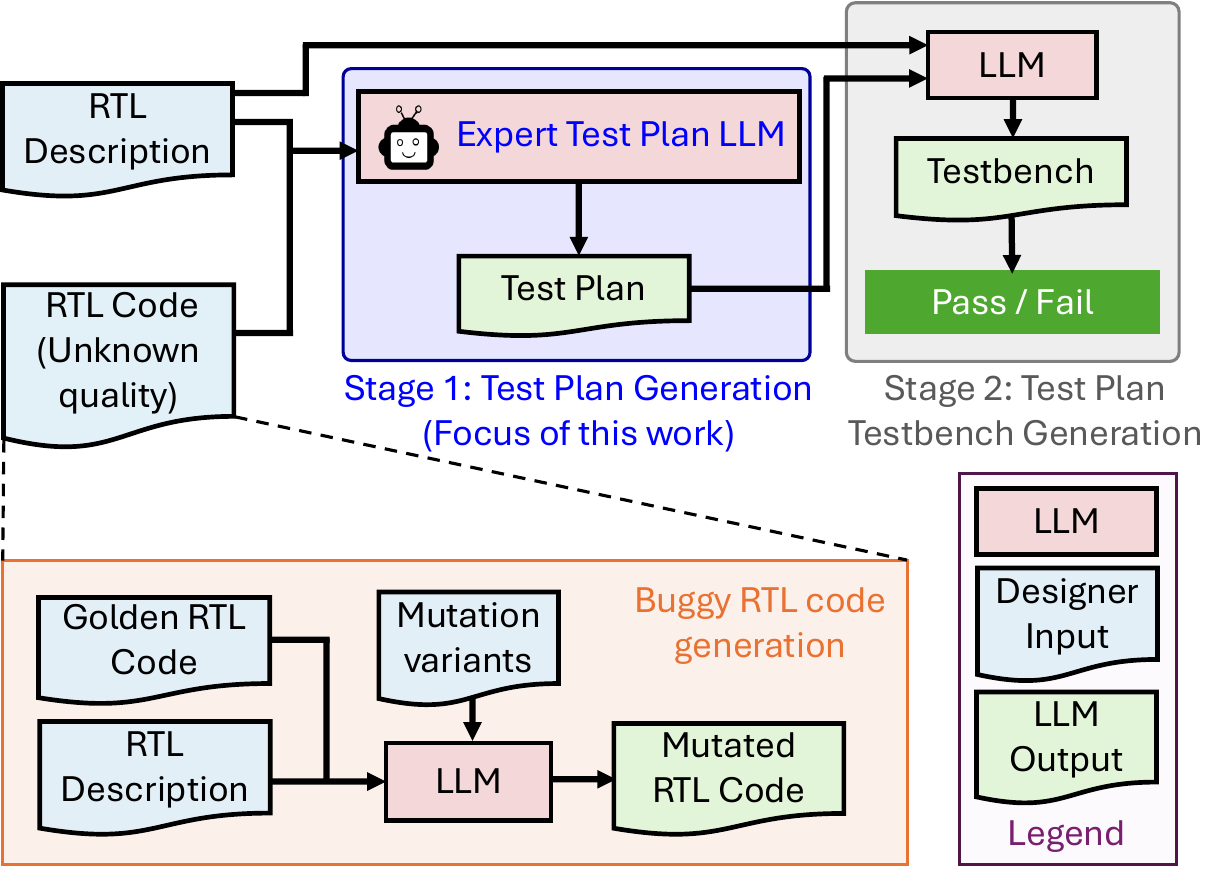}
  \vspace{-0.5em}
  \caption{Our proposed two-stage framework for LLM-based Test Plan Generation and Implementation (top). Buggy RTL code generation for Test Plan Evaluation (bottom), described more in Section~\ref{sec:testdatasetup}.}
  \label{fig:teaser}
  \vspace*{-6mm}
\end{figure}

\dk{To this end, in our work, we tackle the core question: \textit{Can LLMs ``understand'' natural language specifications combined with unknown-quality RTL code to generate high-quality targeted stimuli for debugging?}
This task requires simultaneous reasoning about design intent, implementation details, and potential failure modes.
These capabilities could largely automate and reduce the manual effort required and significantly ease sub-unit verification.}



To investigate this question, we develop a novel two-stage test plan generation framework (as shown in Fig~\ref{fig:teaser}, top).
The first stage is tasked with generating sub-unit test plans \dk{to verify} unknown quality RTL code, while the second stage creates an executable testbench to implement the\dk{se test} plans.
\dk{Our proposed intermediate test plans are verbalized structured representations of LLMs' reasoning for differentiating the RTL description from the RTL design code.
We evaluate \dk{our framework by testing the final testbench} on golden (correct) and mutated (buggy) RTLs, and report the golden pass rate and mutation detection rates as our evaluation metrics.
Overall, our two-stage framework not only improves downstream verification performance by 10-12\% (discussed in Section~\ref{sec:two-stage-ablation}) owing to its task decomposition, but also enhances human control to intervene and improve test plans.}
Benchmarking state-of-the-art LLMs using this approach, we \dk{demonstrate how} the best LLMs achieve a score of $\sim22\%$; revealing significant limitations in LLM reasoning for hardware verification.

Motivated by this limitation, we first explore supervised fine-tuning (SFT) to enhance model reasoning.
Specifically, we fine-tune Deepseek-R1-distilled Qwen-7B LLM on a curated dataset of reasoning traces that demonstrate step-by-step verification planning to distinguish correct and buggy RTL implementations.
However, SFT yields only modest improvements, suggesting that imitation learning alone cannot capture the complex reasoning required for effective test plan generation.

Finally, we apply Group Relative Policy Optimization (GRPO)~\cite{deepseekmath}, a Reinforcement Learning (RL) technique, to teach LLMs directly from verification outcomes rather than mimicking reasoning patterns.
Here, we propose GRPO with State Mutation (GRPO-SMu), which enhances exploration by systematically varying input mutations during training episodes. This approach exposes models to diverse bug patterns within a single training episode. To enable this training approach and generate multiple mutations per RTL, we develop a tree-based code mutation strategy that moves beyond linear mutation approaches.
Our strategy creates functionally equivalent RTL implementations and constructs branching mutation trees that systematically explore bug spaces through validated transformations. Training on this curated data, our 7B LLM achieves a 33.3\% golden pass rate, achieving a $2\times$ improvement over the untrained baseline and outperforming much larger general-purpose LLMs. 

To summarize, we present the following contributions: 
\begin{itemize}
    \item We develop a two-stage framework for systematic benchmarking of LLMs' reasoning capabilities for test plan generation
    \item We propose GRPO with State Mutation (GRPO-SMu) that enhances RL exploration by varying input mutations per specification within a single training episode
    \item We present a novel tree-based mutation strategy for dataset curation that constructs branching equivalent and mutation trees from given RTL code, moving beyond linear mutation approaches
\end{itemize}

\section{Related Work}
\subsection{RTL Code Generation and Verification}
Recent advances in large language models have catalyzed substantial research in automated RTL workflows. Domain-specific fine-tuning approaches~\cite{pearce2020dave,thakur2024verigen,liu2024rtlcoder,pei2024betterv,liu2025deeprtl, scalertl2025} have shown that specialized models can outperform general-purpose alternatives for RTL code generation on Verilog code benchmarks~\cite{liu2023verilogeval, lu2024rtllm}. More sophisticated systems employ multi-agent frameworks~\cite{cui2024origen,gao2024autovcoder,sami2024aivril,mi2024promptv,ho2025verilogcoder}, multi-candidate sampling~\cite{zhao2025vrank,zhao2024mage}, and reinforcement learning~\cite{chen2025chipseek} for enhanced generation quality.

Debugging approaches have also explored retrieval-augmented generation~\cite{tsai2023rtlfixer,yao2024hdldebugger}, iterative refinement~\cite{thakur2023autochip,xu2024meic,huang2024towards}, \dk{stimuli bins for coverage~\cite{zhang2025llm4dv},} and contrastive embedding techniques~\cite{wang2025veridebug}. Automated testbench generation~\cite{ma2024verilogreader,qiu2024autobench,qiu2024correctbench, zhao2025pro} has also emerged as a critical research area. However, these approaches primarily address post-generation correction or testbench automation rather than proactive verification reasoning within the models.

\subsection{Dataset Curation and Mutation Testing}
Prior works like VeriDebug~\cite{wang2025veridebug} insert bugs to build datasets but do not validate functional differences, nor provide multiple variants per RTL code. BugGen~\cite{jasper2025buggen} uses agentic strategies for module-based bug injection but covers only a few designs. Our work is not only more comprehensive, but also addresses data contamination concerns~\cite{vericontaminated, scalertl2025}, as proprietary models likely contain benchmark-related information~\cite{liu2023verilogeval, lu2024rtllm} due to training data cutoff dates.

\subsection{Reinforcement Learning for Language Models}
Supervised fine-tuning provides foundational reasoning capabilities but shows limitations for complex decision-making tasks. Recent reasoning systems indicate that SFT stabilizes models before policy optimization, mitigating failure modes such as looping or incoherent reasoning chains~\cite{deepseek_r1_2025}. Group Relative Policy Optimization (GRPO)~\cite{deepseekmath} samples multiple outputs using average reward as baseline, eliminating the need for separate value networks. GRPO has demonstrated effectiveness in prolonged reasoning tasks~\cite{liu2025prorl} and recently shown promising results in Verilog code generation~\cite{chen2025chipseek}.

\begin{figure}[t]
  \centering
  \includegraphics[width=\linewidth]{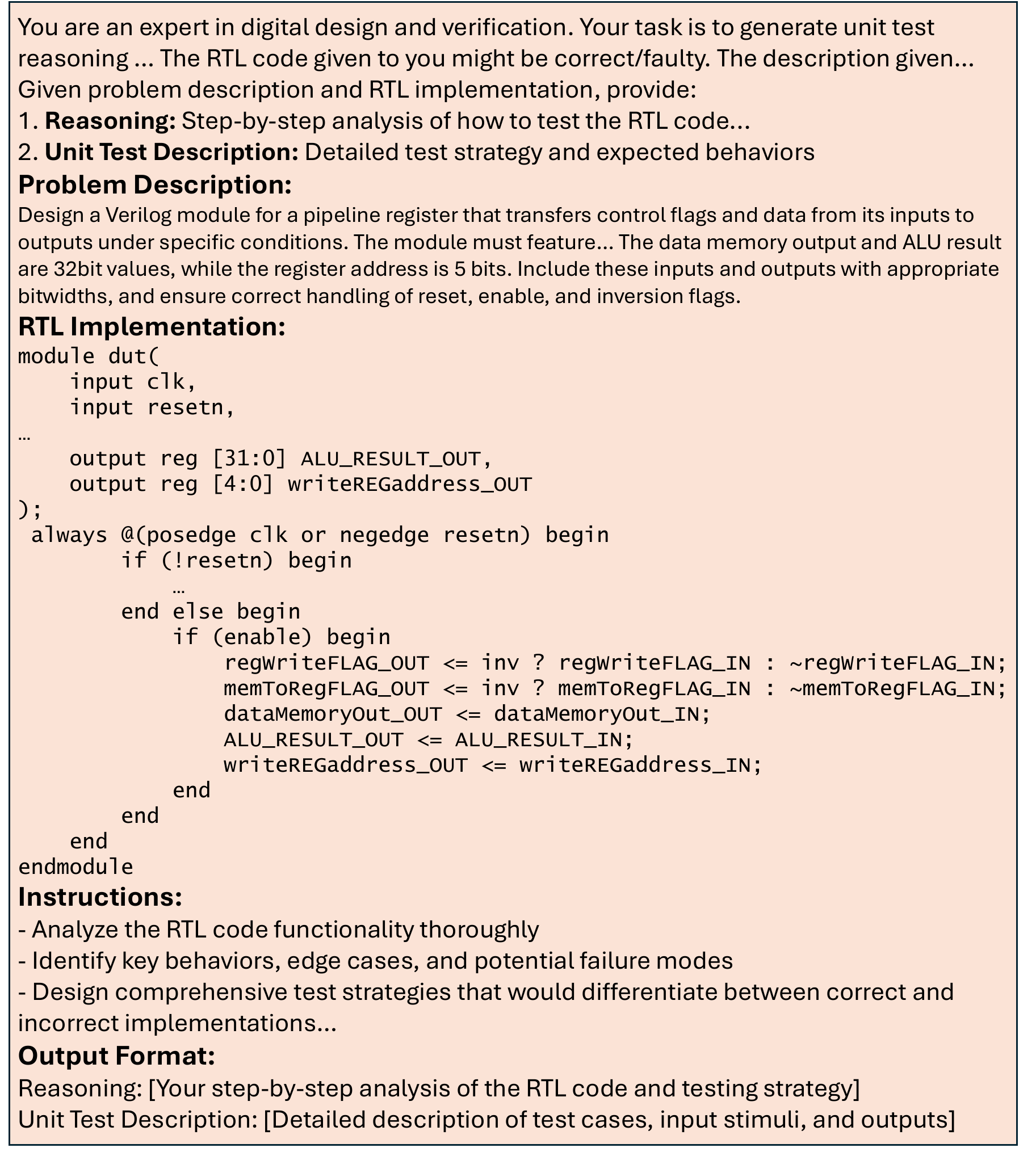}
  \caption{Sample LLM prompt for stage 1, test plan generation, as described in Sec. \ref{sec:test-plan-generation}}
  \label{fig:stage1p}
\end{figure}

\begin{figure}[t]
  \centering
  \includegraphics[width=\linewidth]{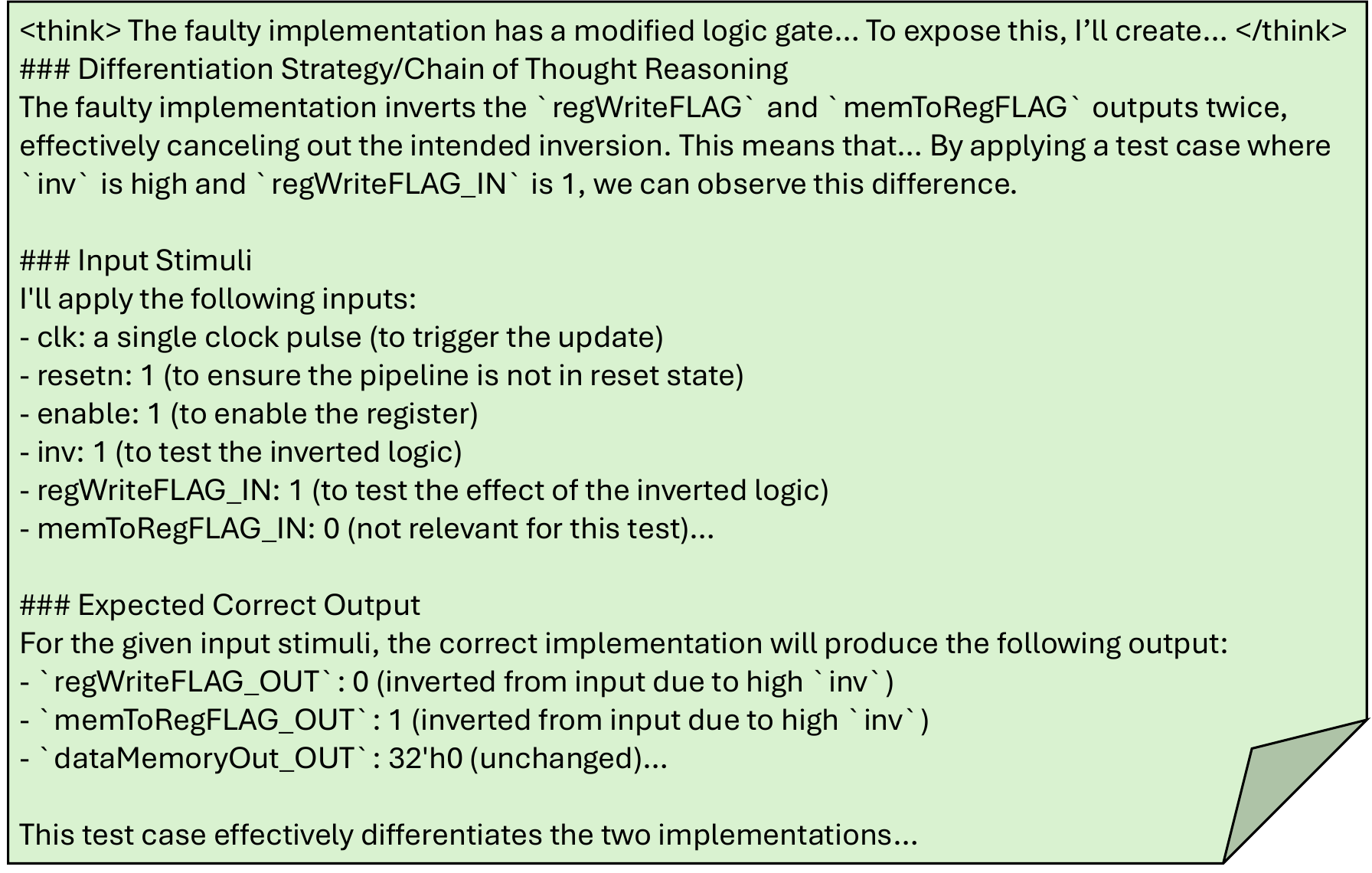}
  \caption{Sample LLM response for stage 1, test plan generation, as described in Sec. \ref{sec:test-plan-generation}}
  \label{fig:stage1r}
\end{figure}

\begin{figure}[t]
  \centering
  \includegraphics[width=\linewidth]{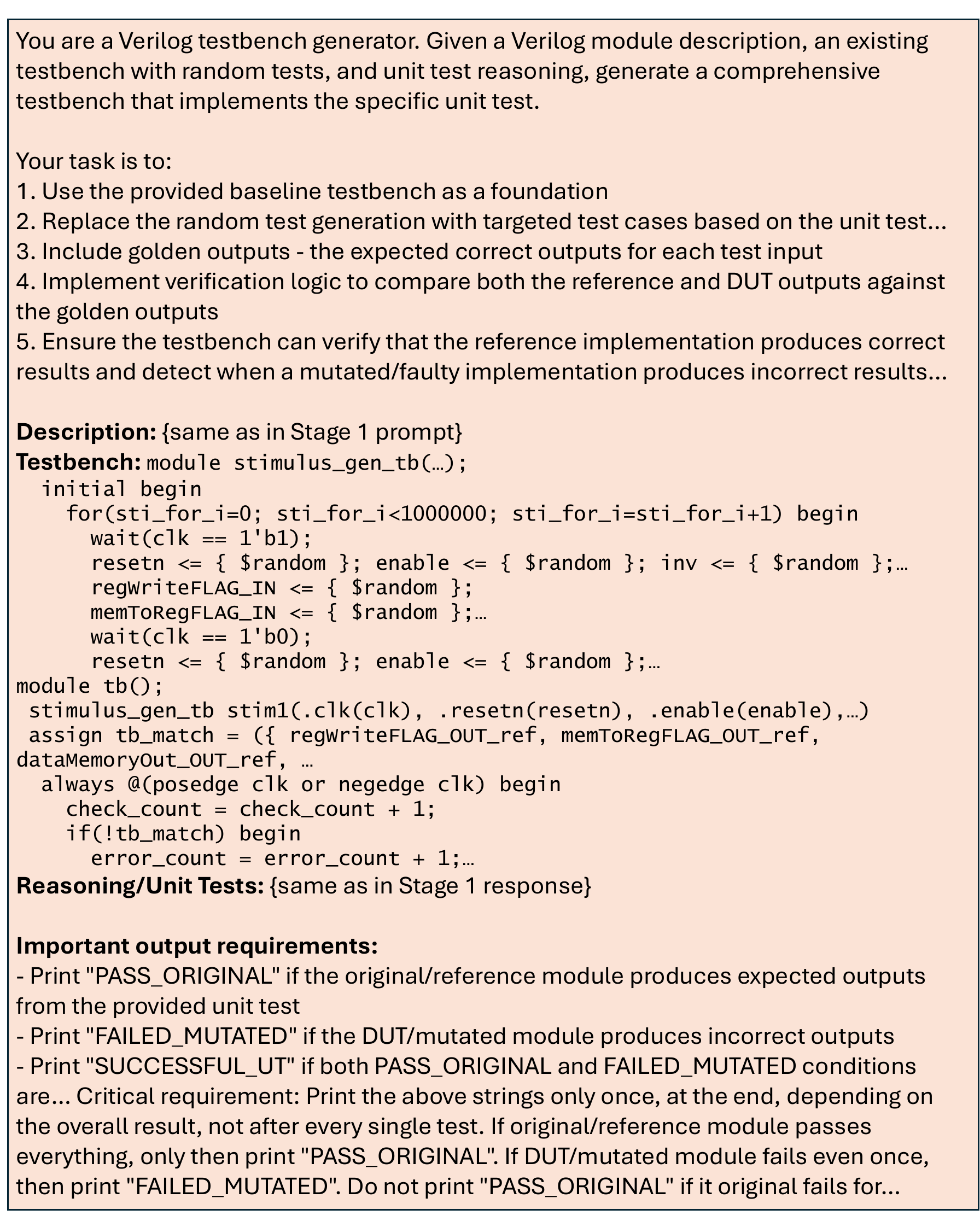}
  \caption{Sample LLM prompt for stage 2, test plan execution, as described in Sec. \ref{sec:test-plan-execution}}
  \label{fig:stage2p}
  \vspace{-.4cm}
\end{figure}

\begin{figure}[t]
  \centering
  \includegraphics[width=\linewidth]{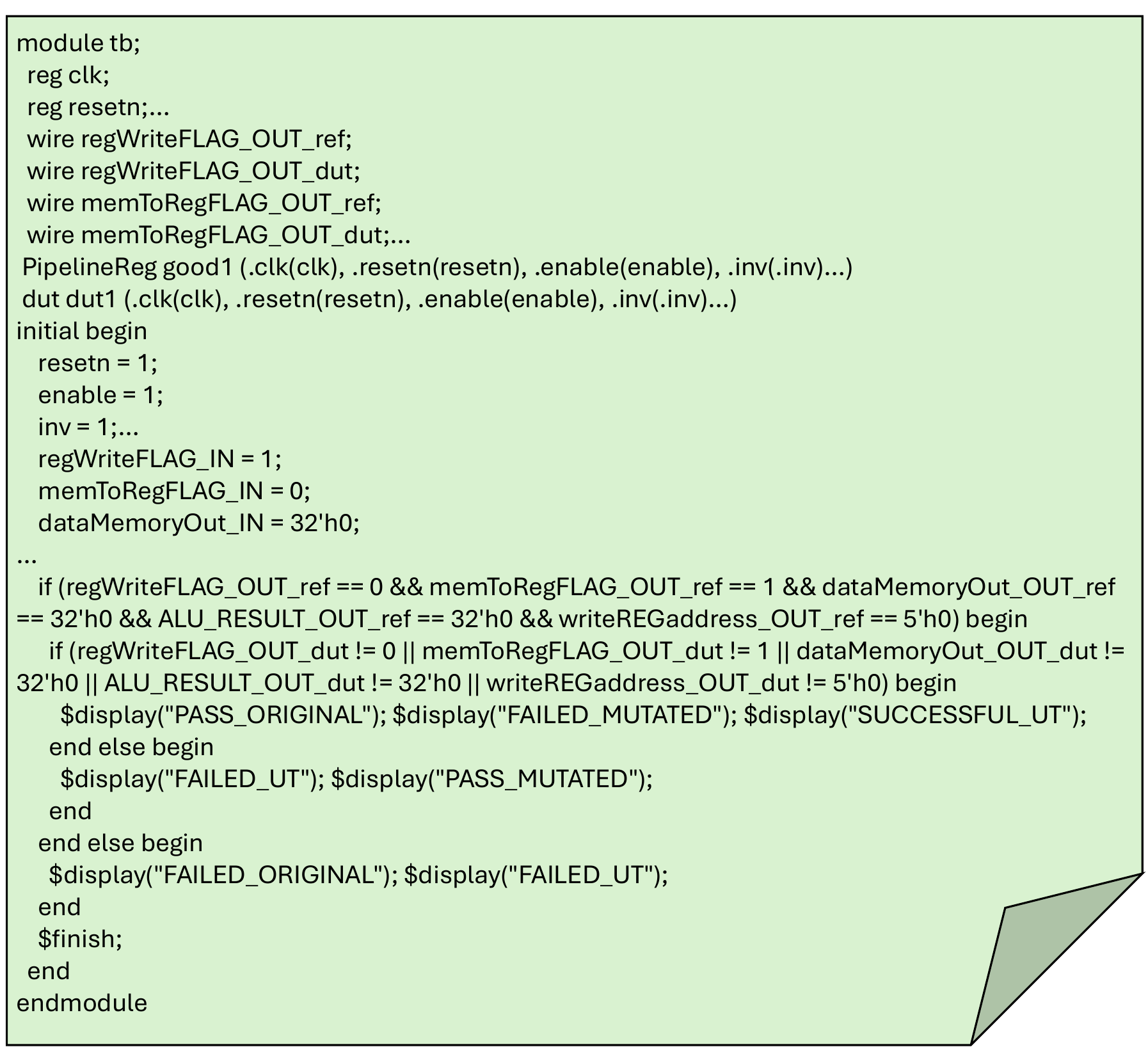}
  \caption{Sample LLM prompt for stage 2, test plan execution, as described in Sec. \ref{sec:test-plan-execution}}
  \label{fig:stage2r}
  \vspace{-.4cm}
\end{figure}
\dk{\section{Problem Formulation: Test Plan Generation}}
\label{sec:start}

\dk{The focus of our work is to develop effective automated approaches to aid low-level RTL design verification, where golden reference models are not available. 
We formulate this task via a two-stage test plan generation framework by introducing an intermediate representation of a test plan.
Here, we describe our problem formulation in depth.}

\dk{\subsection{Two-staged Framework Formulation}
Fig.~\ref{fig:teaser} (top) illustrates our proposed two-stage test plan generation pipeline, introducing a new decomposition to improve performance.
In the first stage, given a natural language RTL description and an RTL design that may contain bugs, the model is tasked to generate a test plan (described in Section~\ref{sec:test-plan-generation}).
In the second stage, the model generates the testbench to implement the test plan from the first stage.
We run this final testbench on correct/golden and buggy/mutated RTL codes (detailed in Section~\ref{sec:testdatasetup}).
Based on this testbench output, we evaluate and report two key metrics for the entire pipeline: (M1) Golden pass rate: whether the golden design correctly passes, and (M2) Mutation detection rate: whether the mutated design is correctly discriminated from the golden design, with the golden design passing.
}

\dk{\subsection{Stage 1: Test Plan Generation:}
\label{sec:test-plan-generation}
We define a test plan as the intermediate reasoning evaluation proxy that distinguishes the golden (reference) behavior from any buggy RTL implementation.
This representation mimics the thinking of a human verifier when developing manual unit tests.
In our work, we formally verbalize it using a structured form comprising (1) the difference between the desired design description and the implemented RTL code, (2) the input stimuli that can elicit this difference, (3) the expected output for verification, and (4) the supporting reasoning. We provide an illustration of a test plan in Fig.~\ref{fig:stage1r}.
}

\begin{figure*}[!t]
    \centering
    \includegraphics[width=0.98\textwidth]{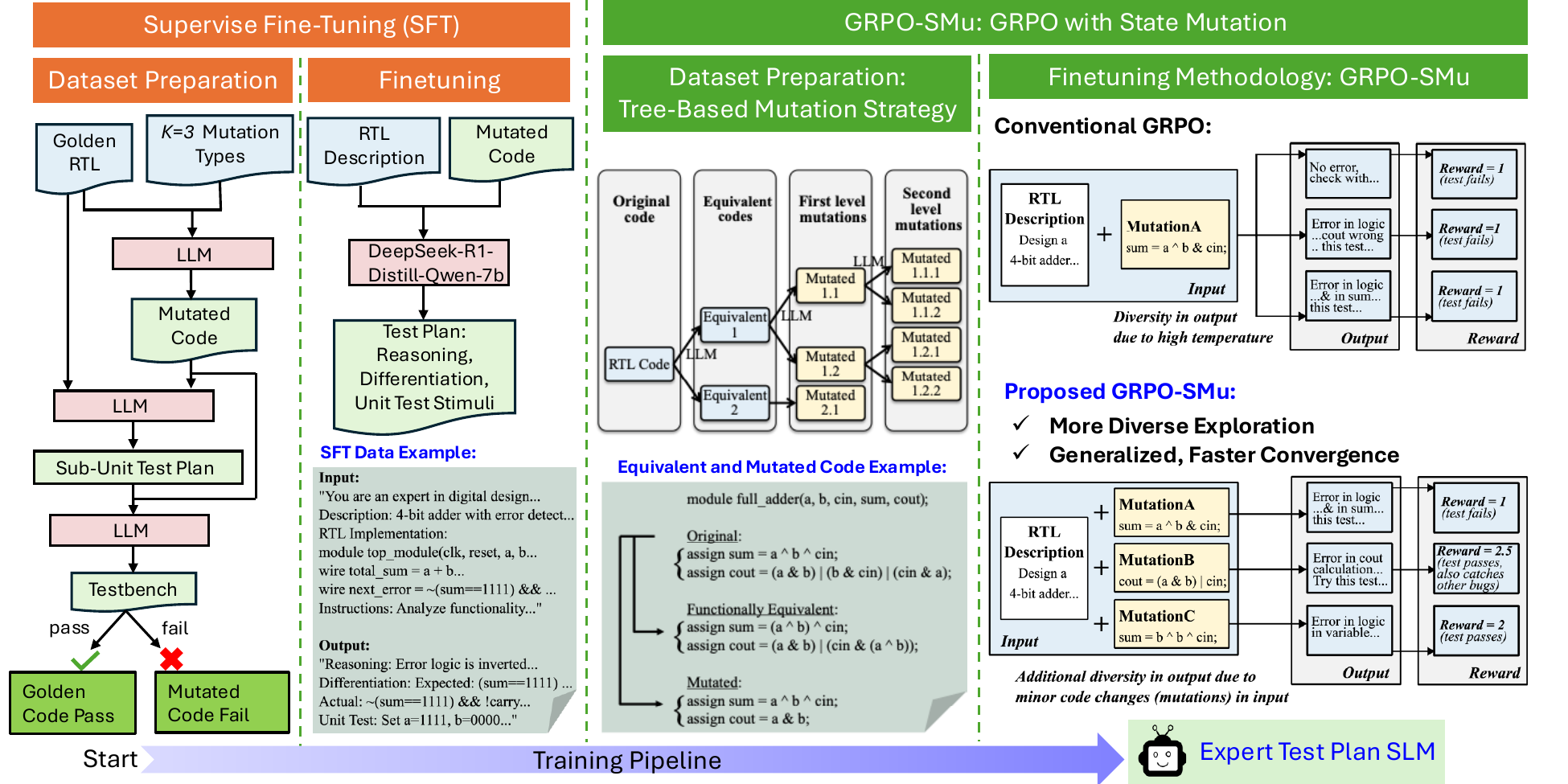}
    \vspace*{-.2cm}
    \caption{Overall training flow: SFT with dataset curation (left) and SFT with example (centre left); Dataset for RL fine-tuning (centre right) and GRPO-SMu vs conventional GRPO comparison (right) showing diverse exploration.}
    \label{overall}
    \vspace{-12pt}
\end{figure*}

\dk{This stage of test plan generation expects the LLM model to understand the natural language specifications, assess the provided RTL code of unknown quality, and generate the correct stimuli through the test plan to expose bugs.
Generating this intermediate test plan provides two major benefits:
(1) Better performance: Similar to previous task decomposition works \cite{khot2022decomposed}, we show performance improvements of 10-12\% (Section~\ref{sec:two-stage-ablation}) using our two-staged task decomposition, compared to single-stage testbench generation, and
(2) Enhanced control: The intermediate test plan provides more control to the human verifier to edit and improve the test plan, compared to the single-stage black-box testbench generation.
We majorly only focus on improving LLMs for this first stage of test plan generation in our work and demonstrate the LLM prompt for this stage in Fig.~\ref{fig:stage1p}.
}

\dk{\subsection{Stage 2: Test Plan Execution}
\label{sec:test-plan-execution}
To execute the first-stage generated test plan logic, a large LLM (we use LLaMa-3.1-405B) processes the generated test plan, combined with a testbench template, to generate a testbench during the second stage.
The testbench template is simply an automatically-generated skeleton testbench stub that helps avoid syntax errors during this testbench generation.
We illustrate the LLM prompt and response for this stage in Fig.~\ref{fig:stage2p} and Fig.~\ref{fig:stage2r}.
}

\subsection{Evaluation Data Creation}
\label{sec:testdatasetup}
For our evaluation, we utilize 500 RTL codes sourced from the ScaleRTL dataset~\cite{scalertl2025}.
To conduct a thorough study, we prompt an LLM to generate \dk{realistic} mutated variants \dk{\cite{jasper2025buggen}} for each RTL code using one of our 14 self-defined bug types (detailed in Section~\ref{sec:data-for-rl}). We illustrate this process in Fig~\ref{fig:teaser} (bottom).
This process creates three mutation variants per code, yielding a final evaluation dataset comprising 1,500 mutated RTL codes. This mutated variant generation addresses data contamination concerns~\cite{vericontaminated}, with proprietary models likely containing previous benchmark information~\cite{liu2023verilogeval, lu2024rtllm} due to training data cutoff dates after their public release~\cite{scalertl2025}.
\section{Methodology}
\dk{For the first stage of test plan generation, we utilize small language models (SLMs).}
\dk{Here, we describe our proposed training methodology to fine-tune these SLMs for our task.}
First, we utilize supervised fine-tuning (SFT) to induce the generation of reasoning traces and structured outputs.
Next, we deploy GRPO-SMu, an RL approach, to fine-tune LLMs using signals from the actual M1/M2 outcomes.
Finally, we discuss our novel mutation generation strategy to curate high-quality data for GRPO-SMu.
We describe these stages and other technical details of our methodology below.

\subsection{Model Preparation: Supervised Fine-Tuning (SFT)}

Fig.~\ref{overall} (left) illustrates our SFT data curation pipeline. We construct SFT pairs from RTL designs drawn from the ScaleRTL dataset by injecting three independent mutations (Table~\ref{tab:mutation_cats}) into each code instance. This approach prioritizes high-yield, discriminative supervision over distributional coverage, increasing the probability that stimuli distinguish golden from mutated code. The pipeline operates as follows: starting with a single RTL design, we generate mutated code, prompt an LLM to create a stimulus test plan to differentiate between correct and mutated code, prompt an LLM to create a testbench from that plan, and simulate it. We accept examples only if the M2 condition holds true (i.e., the golden code passes and the mutated code is correctly discriminated). Using LLaMa-3.1-405B and Claude3.7-Sonnet, we curated 1,902 SFT instances. Each accepted input-output trace is augmented with reasoning traces generated by Claude3.7-Sonnet, yielding (prompt, reasoning + test plan) training data pairs.
Finally, we train the SLM using our SFT dataset, split into 1,521 training and 381 validation samples.
Fig.~\ref{overall} (centre left) illustrates the input-output structure of our training samples with an example.

\subsection{Reinforcement Learning Fine-tuning (RL): GRPO-SMu}
Our work modifies and improves the RL fine-tuning method of Group Relative Policy Optimization (GRPO)~\cite{deepseekmath}.
In GRPO, each training episode comprises a single input state $s$ (i.e., the same initial prompt), from which multiple actions $a_1, a_2, ..., a_n$ are sampled (i.e., different outputs based on temperature). Each action $a_i$ is rewarded with a reward $r_i$ using an external reward model.
The reward distribution creates preference signals (advantages) that are used to train the original action sampling LLM.
However, when the task is complex and the reward signals are sparse (as is the case for our task setting), the reward distribution can lack variation, i.e., all rewards can be equally poor.
The advantages approach zero in such cases, resulting in minimal learning signals~\cite{dapo}.

To address this limitation, we introduce our new policy optimization method: GRPO with State Mutation (GRPO-SMu). Primarily, instead of sampling multiple actions from a fixed state, we sample actions from a systematically diversified set of states, achieving greater action and reward diversity, and in turn, improving the LLM training signal.
We provide more technical details below.

\textbf{Mathematical Formulation: }
In conventional GRPO (Fig~\ref{overall}, right), for each state $s$, multiple actions $a_1, \dots, a_G$ are sampled from the policy $\pi_{\theta_t}$.
For our task, the state comprises the RTL description and the mutated RTL code, the policy model is the LLM, and the actions are the output test plans.
Based on the rewards $r(s, a_j)$ from our reward model (detailed in Section~\ref{sec:reward}), group-relative advantages are computed as:

\begin{equation}
A^{\pi_{\theta_t}}(s, a_{j}) = \frac{r(s, a_{j}) - \mu}{\sigma}
\label{eq:adv}
\end{equation}

where $\mu, \sigma$ are the mean and standard deviation of the grouped rewards. When all rewards are the same, the above advantages will be zero, and there will be no training signal.
This can particularly happen for difficult samples where the action diversity is restricted, leading to restricted exploration.

To overcome this limitation and encourage better exploration, our GRPO-SMu approach (Fig~\ref{overall} (right)) diversifies the initial state space by introducing multiple related states as:
\begin{equation}
s' = s + \Delta s_{\text{mutation}}
\end{equation}
where $\Delta s_{\text{mutation}_i}$ represents different mutation variants for the same RTL code.
Since the new diversified states only differ slightly (they share the same description and the base RTL code from which mutations are created), the states remain sufficiently similar for meaningful advantage computation (Eq~\ref{eq:adv}).
However, sampling actions from this new diversified set of states $\mathcal{S}' = \{s'_1, s'_2, \dots\}$ creates better diversified actions.
The new advantages are computed as:
\begin{equation}
A^{\pi_{\theta_t}}(s'_j, a_{j}) = \frac{r(s'_j, a_{j}) - \mu_{\{\mathcal{S}'\}}}{\sigma_{\{\mathcal{S}'\}}}
\end{equation}
where $\mu_{\{\mathcal{S}'\}}$ and $\sigma_{\{\mathcal{S}'\}}$ are computed from rewards across the set of diversified states $\mathcal{S}'$.
This diversification leads to better exploration and, in turn, increases the variance in reward distribution $\sigma_{\{\mathcal{S}'\}}$ because different mutations $\Delta s_{\text{mutation}_i}$ expose distinct bug patterns.
Eventually, this would lead to more informative advantage signals that capture varied aspects of model performance and, in turn, more efficient and effective LLM training.

The final GRPO-SMu objective function for a given training sample is:
\begin{align*}
L(s'_i, a_i, \theta_t, \theta) &=  \text{clip}\left(\frac{\pi_\theta(a_i|s_i)}{\pi_{\theta_t}(a_i|s_i)}, A^{\pi_{\theta_t}}(s'_i, a_i)\right) \\
& \quad \quad - \beta D_{KL}(\pi_{\theta_t} || \pi_{\theta})
\end{align*}
where $D_{KL}$ is the KL-divergence loss, and the clipping function is defined as:
$$\text{clip}(r, A) = \begin{cases}
\min(r, 1+\epsilon) \cdot A & \text{if } A > 0 \\
\max(r, 1-\epsilon) \cdot A & \text{if } A \leq 0
\end{cases}$$



\subsection{RL Training Dataset: Tree-based Mutation Strategy}
\label{sec:data-for-rl}
GRPO-SMu necessitates multiple mutation variants per RTL code to create the diversified states.
To enable this, we propose a tree-based approach that generates more comprehensive mutations than existing dataset curation methods~\cite{wang2025veridebug, jasper2025buggen}.

Fig.~\ref{overall} (centre right) illustrates our branching tree construction strategy. Starting from golden RTL codes sourced from the ScaleRTL dataset~\cite{scalertl2025}, we generate $n=5$ functionally equivalent RTL implementations (Table~\ref{tab:eq_mut_cats}) as multiple roots for each specification. For each root, we introduce $n_1=3$ first-level mutations drawn from 14 high-level bug categories (Table~\ref{tab:mutation_cats}). We then continue mutating each buggy variant using an expanded operator set of 71 fine-grained operators, constructed from failure analysis of the CVDP benchmark~\cite{pinckney2025cvdp}, yielding deeper branches. This converts a linear process (golden → few random bugs) into a branching tree:
golden → $n$ equivalents → $n \times n_1$ first-level bugs → and so on $\dots$, improving the coverage of the mutation space.


\begin{table}[t]
\centering
\scriptsize
\setlength{\tabcolsep}{3pt}
\caption{Categories for functionally equivalent code generation}
\begin{tabular}{@{}lp{5.8cm}@{}}
\toprule
\multicolumn{2}{c}{\textbf{Functionally equivalent code categories}}\\
\midrule
Stylistic & Different naming, formatting, statement reorganization\\
Architectural & Behavioral vs.\ structural, FSM encodings, decomposition\\
Implementation & Always/case/if restructuring, blocking choices, refactors\\
Optimization & Resource/timing trade-offs, strength reduction\\
Structural & Module hierarchy, signal factoring, layout reorganization\\
\bottomrule
\end{tabular}
\vspace{-.4cm}
\label{tab:eq_mut_cats}
\end{table}

\begin{table}[t]
\centering
\scriptsize
\setlength{\tabcolsep}{3pt}
\caption{Mutation types for branching tree construction}
\begin{tabular}{@{}lll@{}}
\toprule
\multicolumn{3}{c}{\textbf{First-level mutation categories}}\\
\midrule
logic\_gate\_swap & always\_block\_type & blocking\_nonblocking\\
signal\_inversion & condition\_boundary & off\_by\_one\_indexing\\
operator\_swap & timing\_construct & reset\_logic\_error\\
sensitivity\_list & race\_condition & fsm\_state\_corruption\\
 bit\_width\_mismatch & variable\_swap\_within\_line & \\
\bottomrule
\end{tabular}
\vspace{-.4cm}
\label{tab:mutation_cats}
\end{table}

\paragraph{Functional validation}
All equivalence variants and mutations are validated through a random test generator that produces 1M test cases\footnote{\dk{We use this random test generator only for verifying mutation effectiveness in training samples and it does not affect LLM's inference latency.}} per RTL design. For each candidate (equivalence variant or mutation), we run these test cases on both golden (reference) and candidate implementations, counting mismatches in observable outputs. Based on the mismatch statistics: if there are no mismatches, the mutation has no functional effect and is added to \texttt{clean\_codes}; otherwise, the mutation is functionally different and added to \texttt{mutated\_codes}. This functional equivalence checking approach ensures that all transformations are functionally validated rather than relying on purely syntactic criteria.

\begin{figure*}[t]
  \centering
  \includegraphics[width=0.92\linewidth]{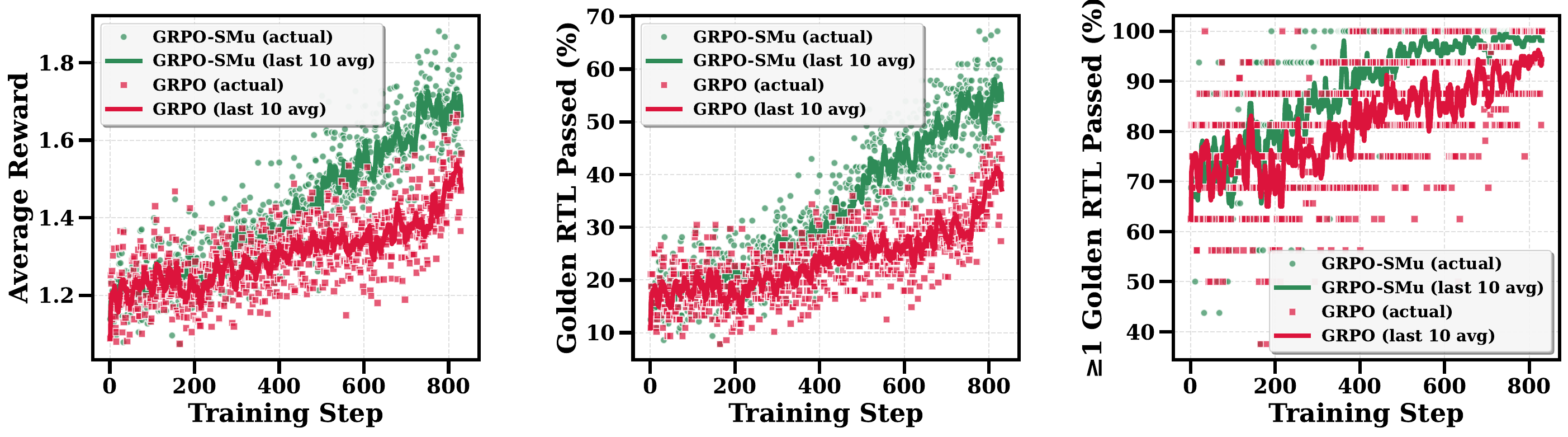}
  \vspace{-.3cm}
 \caption{(a) Training reward comparison between conventional GRPO and GRPO-SMu showing 10\% improvement in training performance over equivalent step count, (b) improved golden pass rate, (c) higher percentage of samples having at least one generation succeed.}
  \label{fig:fig_cross_grpo_grpo}
  \vspace{-.5cm}
\end{figure*}

\paragraph{Data Statistics}
Our validated corpus contains: Base dataset = 2,452 RTL codes; Equivalent codes + Base codes = 14,061 codes; Single-level mutations = 17,926 codes; 1,204 samples selected for level-2 mutations; 22,051 double-level mutations. For GRPO-SMu training, we used 1,318 samples, while we shortlisted 500 samples for the final evaluation.

\subsection{Reward Modeling}
\label{sec:reward}
To distinguish test plans that capture more mutations, we create a nuanced reward model operating on a 0-3 scale:
\begin{equation}
R = r_o + w_m r_m + r_j + r_c
\end{equation}
where:
\begin{itemize}
\item $r_o \in \{0,1\}$ measures the basic functionality (M1: golden code pass)
\item $r_m \in [0,1]$ measures bug detection capability (M2: mutation failure rate), where $r_m = \frac{\#\text{muts\_failed}}{\#\text{total\_muts}}$
\item $r_j \in \{0,0.8\}$ represents external LLM quality assessment
\item $r_c \in \{0,0.2\}$ prevents non-English character generation
\end{itemize}
The weight $w_m$ is set to $1$ only when $r_o = 1$, else $w_m = 0$, ensuring mutation detection rewards are granted only when basic functionality is achieved, to prevent reward hacking.

\subsection{Training}
Our training pipeline
takes the RTL description and mutated code as input. The SLM, initialized from our SFT checkpoint, generates reasoning and unit test plans that differentiate between golden and mutated implementations. To evaluate test plan quality, generated test plans are combined with testbench modules from our random test generator and used to prompt LLaMa-3.1-405B, which executes tests and reports pass/fail status. Critically, this two-component separation prevents reward hacking: the reasoning model generates test plans without knowing the specific prompts or reward structure used by the evaluation LLM, maintaining output quality.


\subsubsection*{Training Challenges}
One of the biggest challenges in training for our task was the reward sparsity. In fact, $\sim30\%$ of generations in the first 100 training steps received rewards, where only one generation achieved a reward $>1$ while $G-1$ were 1. We navigated this in two ways, described below.

\paragraph{Disabling token-level loss}
Using token-level loss, as proposed in DAPO~\cite{dapo}, would lead the LLM to generate mostly uninformative content while keeping token count low, producing occasional high-reward outputs to maximize average group performance. Turning off token-level loss helped mitigate this issue.

\paragraph{Better Advantage Computation}
Under sparse-reward scenarios, leave-one-out baselines use a standard deviation of 1 for the high-reward sample, resulting in weak learning signals (illustrated in Table~\ref{tab:sparse_reward_fix}). To address such cases, we modified the advantage calculation by setting the standard deviation of the entire population (and not leave-one-out). This amplifies the training signal by more than $2\times$.

\begin{table}[t]
\centering
\scriptsize
\setlength{\tabcolsep}{2pt}
\caption{Sparse reward mitigation through modified advantage calculation}
\label{tab:sparse_reward_fix}
\begin{tabular}{@{}p{0.8cm}p{3.0cm}p{3.8cm}@{}}
\toprule
\textbf{Case} & \textbf{Rewards} & \textbf{Standard Advantages} \\
\midrule
Case 1 & $[1.0, 1.0, 1.0, 1.0,$ $1.0, 2.0, 1.0, 1.0]$ & $[-0.38, -0.38, -0.38, -0.38,$ $-0.38, 1.0, -0.38, -0.38]$ \\
\midrule
Case 2 & $[1.0, 1.0, 1.0, 1.0,$ $1.0, 2.0, 0.8, 1.0]$ & $[-0.29, -0.29, -0.29, -0.29,$ $-0.29, 13.61, -0.91, -0.29]$ \\
\bottomrule
\end{tabular}
\begin{flushleft}
\scriptsize
\textbf{Note:} Case 1 represents the sparse reward scenario ($\sim$30\% of training samples) where our modification provides stronger learning signals for high-reward samples. \\Fixed advantage for reward 2.0 in Case 1: $2.83$.
\end{flushleft}
\vspace{-.7cm}
\end{table}

\section{Experiments and Results}
\subsection{Baselines and Implementation Details}
For our baselines, we consider various state-of-the-art general-purpose LLMs like DeepSeek-R1, Claude-4.0-Sonnet, and LLaMa-3.1-405B.
We also use ScaleRTL-32B~\cite{scalertl2025}, an RTL-specific fine-tuned LLM.
For SLMs, we consider the base DeepSeek-R1-distill-Qwen-7B model, the base SFT, and the base GRPO as the baselines. 
The inference temperature was set to 0.7 across all models.
For SFT, we set the learning rate as 2e-6, and global and micro batch sizes as 32 and 2, respectively. 
For GRPO, we use a batch size of 64, with 16 samples per step, 8 generations per sample, a learning rate of 5e-7, a temperature of 1, and a KL coefficient of 0.01.

\subsection{GRPO vs GRPO-SMu Comparison}
First, we directly compare our proposed GRPO-SMu method with the conventional GRPO.
Fig.~\ref{fig:fig_cross_grpo_grpo}(a) shows how GRPO-SMu achieves 10\% higher rewards, on average, compared to GRPO.
These improved rewards are also supported by improved golden pass rate ($\sim$15\% more) as shown in Fig.~\ref{fig:fig_cross_grpo_grpo}(b).
Finally, in Fig.~\ref{fig:fig_cross_grpo_grpo}(c), we show how GRPO-SMu leads to a higher percentage of training groups with at least one test plan succeeding of the 8 generations on the golden RTL, which translates to more reward diversity and better exploration. 

\subsection{Main Results}
Here, we provide the main benchmarking results of all models on our main test setup of the 1500 test samples (Sections \ref{sec:testdatasetup}).
Specifically, given the mutated RTL code and the natural language description, the LLM is tasked with generating the unit test plans.
These test plans are then processed by LLaMa-3.1-405B to generate testbenches.
Finally, we report two key metrics: golden RTL pass and mutation detection rates.
We report our results in Table~\ref{tab:model_comparison}.



\begin{table}[h!]
\centering
\setlength{\tabcolsep}{3pt}
\caption{Model Performance Comparison on Generating Golden Tests and Mutated Code Detection}
\label{tab:model_comparison}
\begin{tabular}{lcccc}
\toprule
\textbf{Model} & \textbf{Golden} & \textbf{Mutated} & \textbf{Change} \\
 & \textbf{Pass (\%)} & \textbf{Detection (\%)} & \textbf{Score} \\
\midrule
\multicolumn{4}{c}{\textit{Small Language Models (SLMs)}} \\
DeepSeek-R1-distill-Qwen-7B & 15.7 & 6.7 & \textit{baseline} \\
base-SFT & 18.2 & 7.5 & +2.5 \\
base-GRPO & \underline{27.2} & 10.5 & \underline{+11.5} \\
\textbf{base-GRPO-SMu (ours)} & \textbf{33.3} & \textbf{13.9} & \textbf{+17.6} \\
\midrule
\multicolumn{4}{c}{\textit{Large Language Models (LLMs)}} \\
DeepSeek-R1 & 21.7 & 10.1 & +6.0 \\
ScaleRTL-32B~\cite{scalertl2025} & 21.6 & 8.9 & +5.9\\
Claude-4.0-Sonnet & 20.6 & \underline{10.6} & +4.9\\
LLaMa-3.1-405B & 16.3 & 6.4 & +0.6\\
LLaMa-3.1-405B-with-golden & 20.2 & 8.7 & +4.5\\
\bottomrule
\end{tabular}
\begin{flushleft}
\footnotesize
\textbf{Note:} Change Score shows percentage point improvement over base SLM model, DeepSeek-R1-distill-Qwen-7B, for Golden Passed metric.
\end{flushleft}
\vspace{-.4cm}
\end{table}

Experimental results reveal how our proposed GRPO-SMu method achieves the best 33.3\% golden test pass rate and 13.9\% mutation detection rate.
Relative to the untrained base DeepSeek-R1-distill-Qwen-7B model, GRPO-SMu improves by more than $2\times$ by up to 17.6\% improvement.
Progressively training from SFT to GRPO to GRPO-SMu, we note improvements of 2.5\%, 8\%, and 6\% respectively, highlighting the significance of fine-tuning to improve LLM's reasoning capability for test plan generation.

Among all LLMs, Deepseek-R1 and ScaleRTL-32B perform best with 21.6-21.7\% golden pass rate.
However, the GRPO-SMu 7B model outperforms both these state-of-the-art models by 11-12\% golden pass rate, further demonstrating the efficacy of our specialized training.


\subsection{Ablating two-stage approach}
\label{sec:two-stage-ablation}
To validate our proposed two-stage pipeline, we compare our approach with a single-stage (standalone) baseline for two LLMs: LLaMa-3.1-405B and Claude-4.0-Sonnet. For standalone prompting, we use the same prompt as our two-stage approach but remove references to reasoning and unit test plans.
As shown in Table~\ref{tab:ablation_study},
our two-stage approach consistently outperforms its single-stage counterparts by 10-12\% golden pass rate for both LLMs.
While Claude-4.0-Sonnet shows better performance for the second stage, we selected LLaMa-3.1-405B for Stage-2 due to additional API costs. 
Finally, we also show that our two-stage 7B GRPO-SMu outperforms all these LLM combinations and is the best test plan generation model.
\begin{figure}[t]
  \centering
  \includegraphics[width=0.95\linewidth]{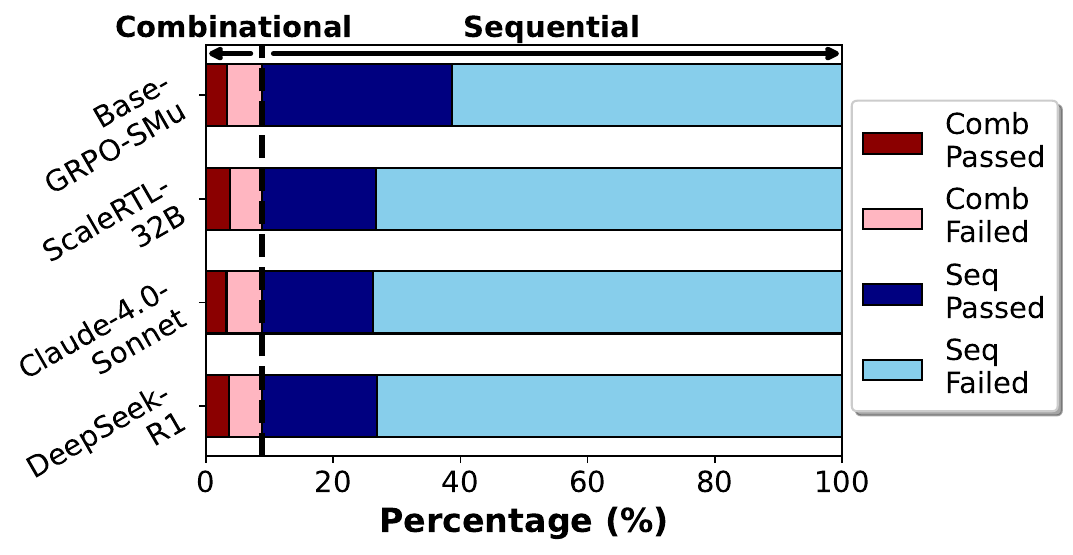}
  \vspace{-0.4cm}
  \caption{Sequential vs Combinational performance improvement study; we see a $\sim1.6\times$ improvement in sequential performance over all general LLMs.}
  \label{fig:comb-seq}
  \vspace{-.4cm}
\end{figure}

\begin{table}[h!]
\centering
\setlength{\tabcolsep}{2pt}
\caption{Two-Stage Architecture Justification}
\label{tab:ablation_study}
\begin{tabular}{lcc}
\toprule
\textbf{Model Configuration} & \textbf{Golden} & \textbf{Improvement} \\
 & \textbf{Pass (\%)} & \textbf{over Baseline} \\
\midrule 
LLaMa-3.1-405B (standalone) & 5.7 & -- \\
LLaMa-3.1-405B $\rightarrow$ LLaMa-3.1-405B & 16.3 & +10.6 \\
Claude-4.0-Sonnet (standalone) & 18.9 & +13.2 \\
Claude-4.0-Sonnet $\rightarrow$ LLaMa-3.1-405B & 20.6 & +14.9 \\
Claude-4.0-Sonnet $\rightarrow$ Claude-4.0-Sonnet & 30.2 & +24.5 \\
\midrule
\textbf{base-GRPO-SMu $\rightarrow$ LLaMa-3.1-405B (ours)} & \textbf{33.3} & \textbf{+27.6} \\
\bottomrule
\end{tabular}
\begin{flushleft}
\footnotesize
\textbf{Note:} All improvements calculated relative to LLaMa-3.1-405B standalone baseline. Two-stage notation: Stage-1 $\rightarrow$ Stage-2.
\end{flushleft}
\vspace{-.4cm}
\end{table}

\subsection{Circuit Type Error Analysis}
We analyze our test benchmark by classifying each sample as combinational or sequential logic using majority voting across 4 LLMs. Of the 500 test samples, 44 (8.8\%) are combinational while others are sequential, validating our focus on sequential debugging. Our training data exhibits a similar distribution. Performance analysis from Fig.~\ref{fig:comb-seq} reveals that all models perform better for the easier combinational circuits.
However, for sequential circuits, we observe nearly 2× improvement by GRPO-SMu compared to all other models---highlighting its source of improvement.

\section{Conclusion and Future Work}

\dk{This work introduces a new task decomposition and proposes a two-stage framework by introducing verbalized test plans as intermediate representations for sub-unit verification.
Such test plans not only improve performance, but also enhance the human-in-the-loop control for building better test plans.}
Our \dk{in-depth evaluation and} investigation reveals that state-of-the-art LLMs achieve only 15-22\% success rates on \dk{our} hardware verification task despite their strong reasoning and code generation capabilities.
To improve \dk{them further}, we propose an enhanced RL technique, GRPO-SMu, that enhances exploration through input mutation diversity.
We also develop a novel tree-based mutation strategy for training data generation for GRPO-SMu.
We demonstrate how our GRPO-SMu \dk{on a small} 7B \dk{LLM} achieves the best 33.3\% golden test pass rate, outperforming other larger and fine-tuned LLMs.  
While LLMs remain far from production-ready verification, our work establishes crucial foundations for autonomous debugging frameworks to replace manual test creation and improve productivity.
Future works can utilize our framework and learnings to develop scalable, deployable systems and agentic workflows, fundamentally reshaping hardware verification to aid faster next-generation semiconductor design.

\bibliography{refs}


\end{document}